\documentclass[prl,showpacs,twocolumn]{revtex4}
\usepackage{amsfonts}
\usepackage{amsmath}
\usepackage{amssymb}
\usepackage{graphicx}

\begin{document}

\title{Conduction Bands of Atomic Tunneling Ring in Artificial Gauge Field Assisted Opened Optical Traps}
\author{Wenxi Lai$^{1}$,Yuquan Ma$^{2}$, and W. M. Liu$^{1}$}\email{wmliu@iphy.ac.cn}
\affiliation{1 Institute of Physics, Chinese Academy of Science, Beijing 100190, China}

\affiliation{2 School of Applied Science, Beijing Information Science and Technology University, Beijing 100192, China}

\begin{abstract}
We show conduction bands of artificial gauge field assisted atom flow in triangle optical lattice. The conduction bands are result from periodicity boundary condition of artificial magnetic flux induced phases of atoms. The positions of conduction bands depend on geometry of the atom trajectory. We consider a cell of the triangle optical lattice which is a opened system connected to its environment of Fermion atom clouds. The chemical potentials of the atom clouds are the same and the atom flow is absolutely created by a clock laser induced spin-orbit coupling. Our results are important for the control of atom flow in quantum circuits.
\end{abstract}

\maketitle

Cold atom systems are very useful for the simulation condensed matter materials and for the design of atomic devices. Recently, artificial gauge field in synthetic dimension of optical lattice brings about many interesting phenomenons. Artificial gauge field for neutral particles can be obtained using coherent light-particle interactions, such as two-photon Raman transition~\cite{Dalibard} or ultra-narrow clock field transitions~\cite{Livi,Kolkowitz}. In optical lattice, the artificial gauge field induces spin-orbit coupling (SOC) of cold atoms, which couples orbital motion of neutral atoms to its internal electronic states. The SOC associated with synthetic space plays an important role in atom transport problems, such as quantum spin Hall effect~\cite{Kennedy}, chiral insulator~\cite{Kollath}, chiral current~\cite{Mancini,An}, superradiance induced particle flow~\cite{Zheng}.

Since in alkali atoms Raman-induced spin flips can not avoid heating mechanisms associated with spontaneous emission, the technique of direct transition between two long-lived electronic clock states has been widely developed recently. Life time of the clock states in alkaline-earth atoms or lanthanide atoms reach from $10$ $s$ to $10^{3}$ $s$~\cite{Nicholson12,Hinkley,Bloom,Nicholson15,Huang,Marti}. In the long coherent atom-light interaction, energy and momentum conservation cause observable non-equilibrium atom flow due to SOC. The narrow linewidth clock transition between a particular ground $|g\rangle$ and excited states $|e\rangle$ would be accompanied by momentum change of atoms with maximum value $2\pi \hbar/\lambda_{C}$ ($\hbar$ is Plank constant, $\lambda_{C}$ wave length of the clock laser). Atom transport effects using the clock transition in synthetic space in one dimensional optical lattice are studied in previous~\cite{Wall,Livi,Zheng,An,Kollath,Recati}. One problem is what would be happened when atom trajectory in the optical lattice is not infinitely long line.

In the letter, we consider tunneling of cold atoms through a few coupled optical traps. In this case, we encounter a boundary condition of the system. the boundary condition is that atoms move a round a closed trajectory in the optical lattice its phase would be integer times of $2 \pi$, and in our system it is always Zero.

\begin{figure}
  \includegraphics[width=8cm]{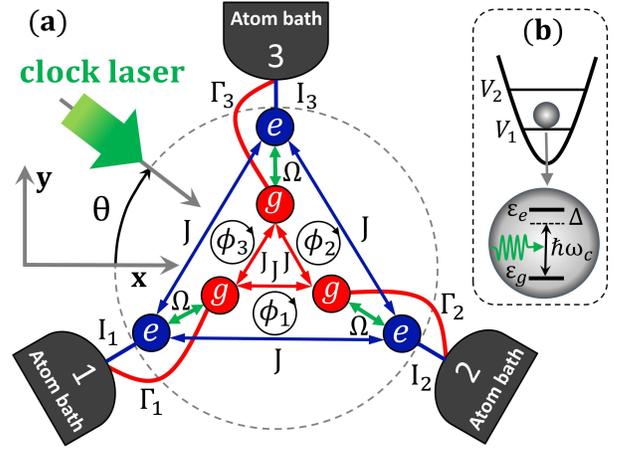}\\
  \caption{(a) (Color on line) The three optical traps form a closed loop. Atoms can move in the loop with clockwise or counterclockwise direction. The atomic loop is characterized by three pole with each one coupled to an atomic lead. An optical clock field acting on the atoms in the optical traps to create artificial magnetic flux in the closed loop of synthetic dimension of the neighboring optical traps. (b) Two energy levels are considered in each optical trap and the clock transitions of atoms are applied here. }\label{sys}
\end{figure}

Our system is illustrated in Fig.~\ref{sys}. Three optical traps are arranged in a triangle geometry and not on a line. Atom tunneling is allowed between neighboring traps~\cite{Recati,Bauer,Knap}, and forms a closed trajectory of the atom transport. This ring is an opened system with each optical trap couples to an atomic bath. The atomic bath can be treated as cold atom clouds or rest part of the atoms in a large triangle optical lattice. Atoms are in or out from the atom bath and initially one can suppose the three optical traps are empty. We consider Fermion cold atom $^{173}Yb$ with the ground state $g=$ $^{1}S_{0}$ and the matastable state $e=$ $^{3}P_{0}$. This two states allow optical clock transition with a coherent life time of $20$ s. Our clock laser is at the wave length  $\lambda_{C}=578$ nm and the laser for optical potential is at wave length $\lambda_{L}=759$ nm. We assume only one atom is allowed to occupy each optical trap at a time due to atom-atom interactions. Therefore, interactions between two atoms are not considered here. In most of the previous artificial gauge field assisted systems, the cites of optical lattice are always arranged in the one line with a definite laser-lattice angle $\theta$ \cite{Livi}. However, in our model the positions of optical lattice cites are not in one line. The atom trajectory is in a two dimensional plane and atom can move along a real closed way. The synthetic dimension of the system that created by the clock laser is in the third dimension of the system(See Fig.~\ref{sys}). However, the artificial magnetic flux would be not linearly changed between the neighboring vortexes, since the angles between clock laser propagation and atom tunneling direction is different for neighboring vortexes.

From atom $1$ to atom $2$, atom $2$ to atom $3$ and atom $3$ to atom $1$, we have the laser angles $\theta_{1}$, $\theta_{2}$ and $\theta_{3}$, where $\theta_{1}=\theta$, $\theta_{2}=\theta+(\pi-\alpha)$ and $\theta_{3}=\theta+(\pi-\alpha)+(\pi-\beta)$ (See Fig.~\ref{sys}), respectively. Then the corresponding synthetic magnetic flux is $\varphi_{1}=\pi \lambda_{L}cos\theta_{1}/\lambda_{C}$, $\varphi_{2}=\pi \lambda_{L}cos\theta_{2}/\lambda_{C}$ and $\varphi_{3}=\pi \lambda_{L}cos\theta_{3}/\lambda_{C}$. The artificial magnetic flux in every closed synthetic trajectory is $\phi_{1}=\phi_{3}+\varphi_{1}$, $\phi_{2}=\phi_{1}+\varphi_{2}$, and $\phi_{3}=\phi_{2}+\varphi_{3}$, respectively. This is the key formulas in the paper which indicate the particular phase relation between neighboring optical traps. The phase difference of the closed tunneling system satisfy a periodicity condition which is different from the phase relations in the infinite long optical lattice chain~\cite{Zheng,An,Kollath,Recati}. In principle, the periodicity condition should be $\phi_{3}=2n\pi$, where $n$ is any integer number. Interestingly, this condition is satisfied naturally in the system with $\phi_{3}=0$.

Between the neighboring two leads, a vortex is created in synthetic dimension of the optical lattice. A clock laser induced synthetic dimension to the atoms trapped in the optical traps. The tunneling between two optical traps and optical coupling of the internal atomic states forms a vortex with effective phase $\phi$. This vortex leads to pure current of the atoms. There are three poles connected to three atomic cloud. Transport properties of the atoms through the tree poles can be controlled by the artificial gauge field and other parameters of the system.

At current operating temperatures, $T\sim \mu K$, the population of higher axial bands is negligible ($\lesssim 5 percent $).

To give a quantitative description to the model, we use the following Hamiltonian

\begin{eqnarray}
H &=& H_{S}+H_{B}.
\label{eq:Hamiltonian}\end{eqnarray}
The first term is Hamiltonian of atoms bounded in the three optical traps,
\begin{eqnarray}
H_{S} &=& \sum_{j,s}\epsilon_{\alpha,s} a_{j,\alpha,s}^{\dag}a_{j,\alpha,s}+\hbar J \sum_{j,\alpha,s} (a_{j,\alpha,s}^{\dag}a_{j+1,\alpha,s}+h.c.)\notag \\
&&+\frac{\hbar \Omega}{2}(e^{i \phi_{j} }e^{i \omega_{c} t} a_{j,\alpha,g}^{\dag}a_{j,\alpha,e}+h.c.),
\end{eqnarray}
Where, energy of these atoms in the optical traps consists of two parts, $\epsilon_{\alpha,s}=V_{\alpha}+E_{s}$ with $\alpha=1, 2$, $s=g, e$. $V_{1}$ and $V_{2}$ are two lowest bounding energy of atoms, they related to the depth of optical potential. $E_{g}$ and $E_{e}$ are energy of electron transition in the atoms that coupled to the clock laser. $a_{j,\alpha,s}$ ($a_{j,\alpha,s}^{\dag}$) is annihilation (creation) operator of atoms in the $j$ optical trap.
The second and third term of $H$ describe tunneling of the atoms between neighboring traps and the clock laser coupling to the atoms. The tunneling amplitudes are described by $t_{j}$. The serial number $j$ satisfies $j+1=1$ when $j=3$. A clock light of frequency $\omega_{c}$ is interacting with the single atoms with the Rabi frequency $\Omega$. The frequency of optical field is very fast at the visible light frequency, therefore we take the rotating wave approximation for the atom-field interactions. The atomic leads are described by the Hamiltonian of free atomic gas
\begin{eqnarray}
H_{B} &=& \sum_{j,s,k}\varepsilon_{s,k} a_{j,s,k}^{\dag}a_{j,s,k} +\sum_{j,k,\alpha,s}(t_{j} a_{j,k,s}^{\dag}a_{j,\alpha,s}+h.c.).\notag \\
\end{eqnarray}

Here, $a_{j,s,k}$ ($a_{j,s,k}^{\dag}$) is annihilation (creation) operator of the atoms in the atom leads $j$. Energy of these atoms can be written as $\varepsilon_{s,k}=V_{k}+E_{s}$, where $V_{k}$ is the energy about atom motion in the potential that bounds the atomic gas. $k$ is wave number of an atom wave function.

Using the Markovian approximation to the coupling between system and atomic clouds, we obtain the following master equation for the atom-light opened system\cite{Wenxi},
\begin{widetext}
\begin{eqnarray}
\frac{\partial}{\partial t}\rho &=&\frac{1}{i} [\frac{\triangle}{\hbar}\sum_{j} a^{\dag}_{j,\alpha,e}a_{j,\alpha,e}+\frac{\Omega}{2} \sum_{j}(e^{i \phi_{j}} a_{j,\alpha,g}^{\dag}a_{j,\alpha,e}+h.c.)+\sum_{j,s} J (a_{j,\alpha,s}^{\dag}a_{j+1,\alpha,s}+h.c.),\rho]\notag \\
&&+\sum_{j,\alpha,s}\frac{\Gamma _{j}}{2}[f_{j}(V_{\alpha})(2 a^{\dag}_{j,\alpha,s}\rho a_{j,\alpha,s}-\{a_{j,\alpha,s}a^{\dag}_{j,\alpha,s},\rho \})+(1-f_{j}(V_{\alpha}))(2 a_{j,\alpha,s}\rho a^{\dag}_{j,\alpha,s}-\{a^{\dag}_{j,\alpha,s}a_{j,\alpha,s},\rho \})]
\label{eq:Hamiltonian}\end{eqnarray}
\end{widetext}
where $\Delta=\varepsilon_{e}-\varepsilon_{g}-\hbar \omega_{c}$ is decoupling of the clock field from the two states of atoms. The Fermi-Dirac distribution function of the Fermi atoms is written as $f_{j}(V_{\alpha})=\frac{1}{e^{(V_{\alpha}-\mu_{j})/k_{B}T}+1}$. Here, $\mu_{j}$ is the chemical potential of the atom lead $j=1,2,3$. Energy of system satisfy $\varepsilon_{e}-\varepsilon_{g}, \hbar \omega_{c}\gg \Omega, J, \Gamma _{j}, \triangle, k_{B}T$.

There is three Fermion atomic leads (denoted by $j=1, 2, 3$, respectively). We set chemical potentials of these atomic leads are the same, $\mu_{1}=\mu_{2}=\mu_{3}$. Therefor it is confirmed that non equilibrium state would not be caused by the chemical potential.

We take atom tunneling rates as $J=2\pi\times 300 Hz$ and optical Rabi frequency as $\Omega=2\pi\times 590 Hz$, $\Delta=2\times J$, $T=0.5 \times J$ and $\Gamma=J$, which are based on the recent experiment \cite{Livi}.

Here, we use polar coordinates and take the direction of clock laser which is acting on the triangular optical lattice as the coordinate axis. In the coordinate system of experiment, we can change the angle of the clock laser in the clockwise direction. However, in the coordinate system of the the clock laser, the atomic optical lattice rotates in the anti-clockwise. We define the rotation direction of the atomic lattice in the coordinate system of the clock laser as $\theta$.

\begin{figure}
\includegraphics[width=0.48\textwidth, clip]{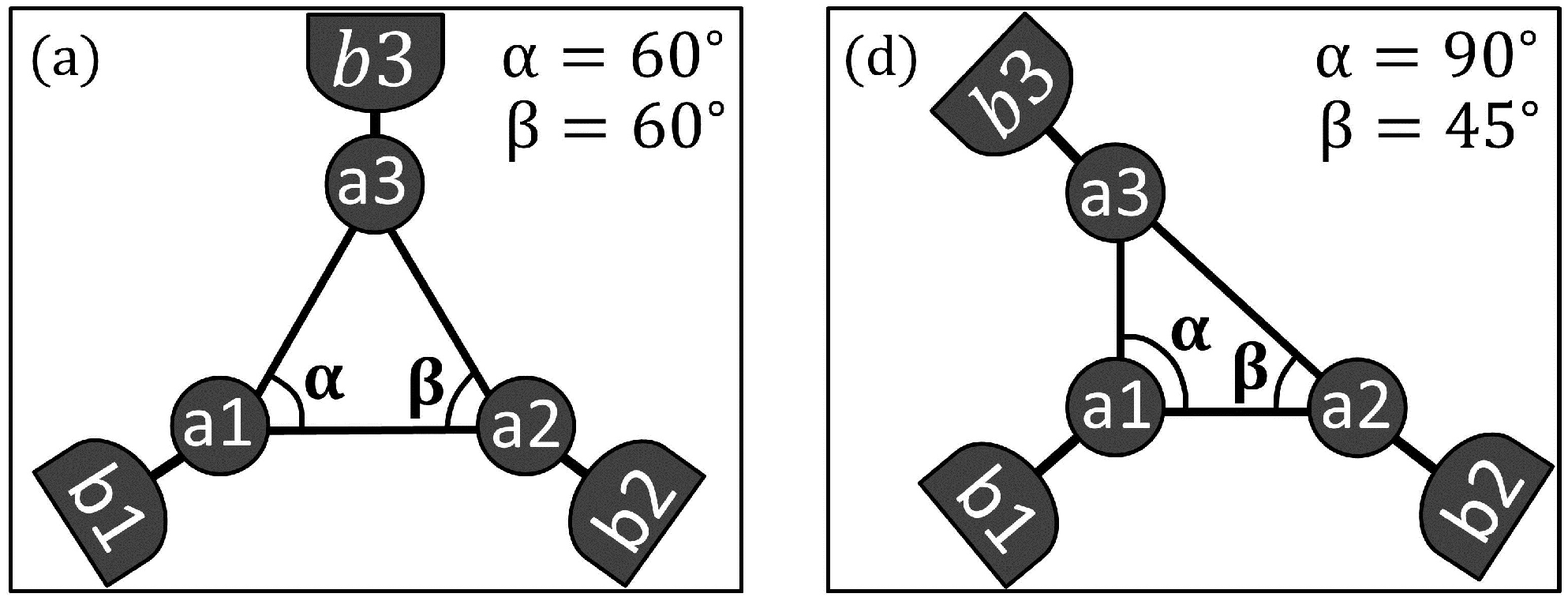}\\
\includegraphics[width=0.48\textwidth, clip]{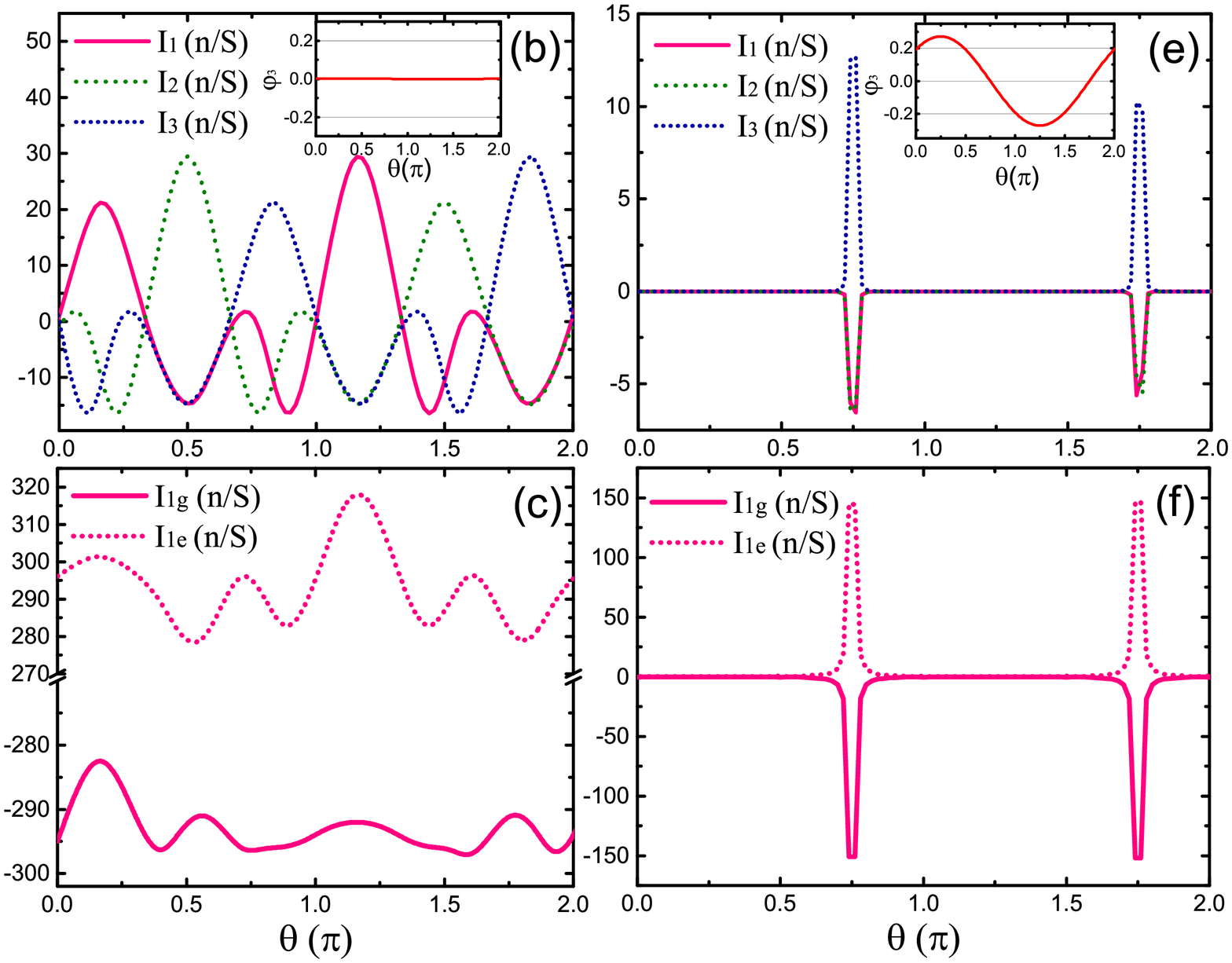}\\
\caption{Atomic current as a function of the laser direction $|\theta|$; (b)Atomic current as a function of the laser
direction $|\theta|$ in the case of one pole of the audion is shut down $(\Gamma_{3}=0)$. Current direction of the
atomic audion corresponding to the laser direction $\theta$ from $0^{\circ}$ to $120^{\circ}$. }%
\label{i-theta}%
\end{figure}

Current of the neutral atomic can be excited by the artificial gauge field that created with the clock laser field on two level atoms\cite{Livi,Mancini,Zheng}. We show that particle flow can also be observed in the closed path of optical lattice. As the most simple model, triangular lattice is considered in our work. Our interest is the three optical traps that forms a smallest unit of the regular triangle lattice. These three optical traps coupled to three atomic leads would consist of a quantum device works like an audion. The system has two rotation symmetry in the plan of the triangle, one is for the rotation of $360^{\circ}$, another is for the rotation of $120^{\circ}$. Therefore, the current configuration of the audion has a period for every $360^{\circ}$ rotation of the clock laser as shown in Fig.~\ref{i-theta}. At the same time, the current configuration of the audion also has three periods in the regime of $0^{\circ}\sim 360^{\circ}$. However, $I_{1}$, $I_{2}$ and $I_{3}$ have different behavior in the three period zone of $0^{\circ}\sim 120^{\circ}$, $120^{\circ}\sim 240^{\circ}$ and $240^{\circ}\sim 360^{\circ}$.

In Fig.~\ref{i-theta}, we illustrate the current states of atoms in the fist period zone $0^{\circ}\sim 120^{\circ}$, denoting the corresponding laser directions. There are altogether six different states which are depend on the angle of clock laser acting on the atomic system. Without considering the direction of the triangle system, we have two kinds of current style in fact, first is one in-pole and two out pole; second two in-pole and one out-pole.

In the triangle configurations, every neighboring pair atoms form a closed circle with the synthetic dimension for artificial gauge field. There for every side has a vertex which cause atom transitions in the circle of the synthetic dimension. And this vortex lead to a particular spin-orbit coupling for each neighboring atoms which is different from other two pair atoms. There are three different spin-orbit coupling for three atom pair. They lead to non-equilibrium atom flow in the closed trajectory of the triangle optical lattice.

The atomic audion is remarkably related to the decoupling of the laser. The current direction would be changed when the decoupling is tuned from red to blue or conversely. For some particular laser direction, such as $\theta=0^{\circ}, 60^{\circ}, 120^{\circ}$, we always have exactly the same current of two pole for any decoupling $\Delta$. Atomic currents for other parameters such as the Rabi frequency $\Omega$, atom tunneling strength $J$ and system-lead coupling strength $\Gamma$ are illustrated, respectively. Since atom current is related phase of the wave functions, it it sensitive to the coherent couplings. When these four parameters are similar, we have remarkable current. However, When one of the parameters is too small or too big, the current would be suppressed.

When we shut one of the three poles of the system, the other two poles work as a diode as shown in Fig.~\ref{i-theta}. The atom flow direction depends on the angle at which the clock laser acting on the atoms.

\begin{figure}
\includegraphics[width=0.24\textwidth, clip]{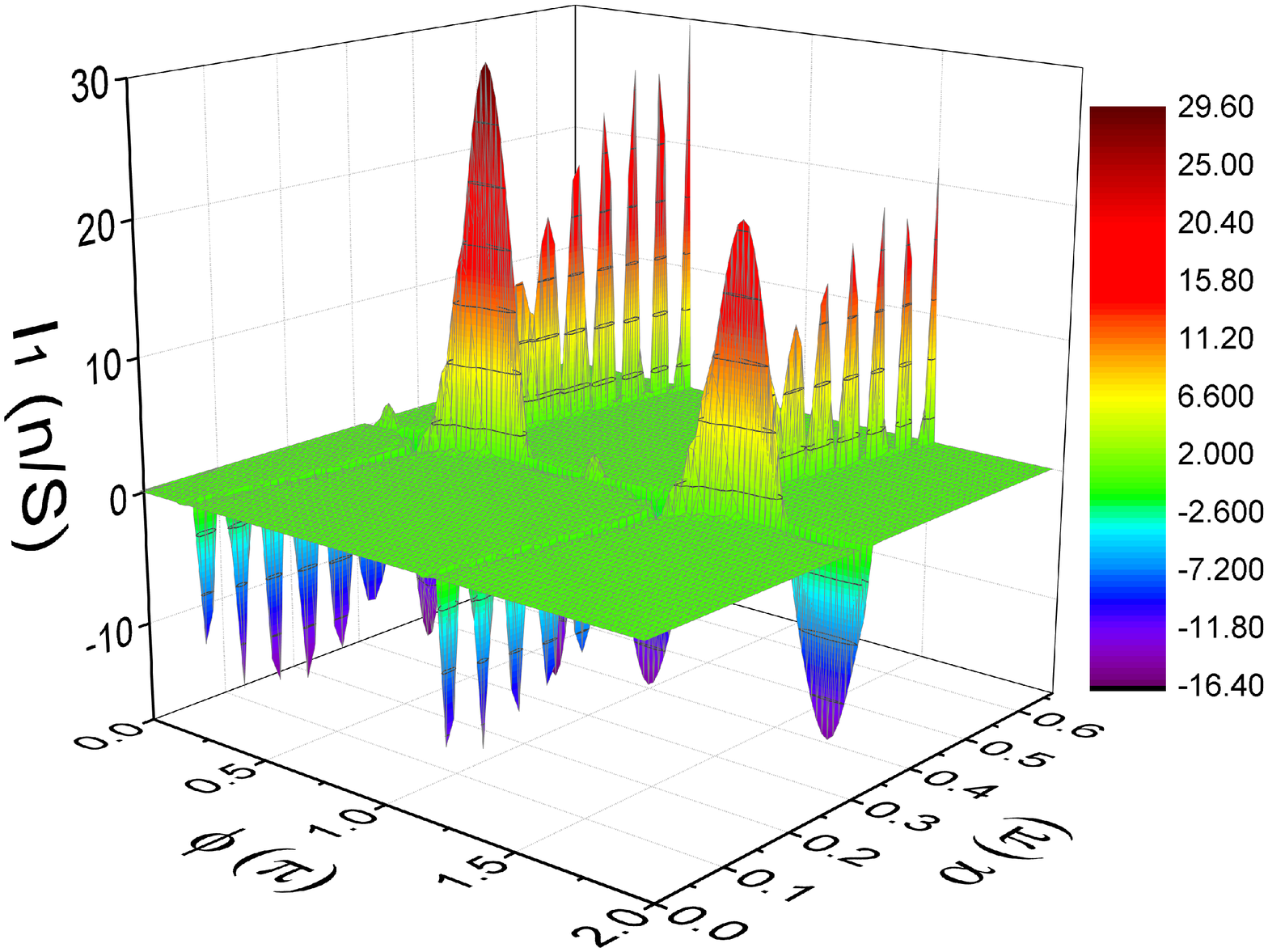}\includegraphics[width=0.24\textwidth, clip]{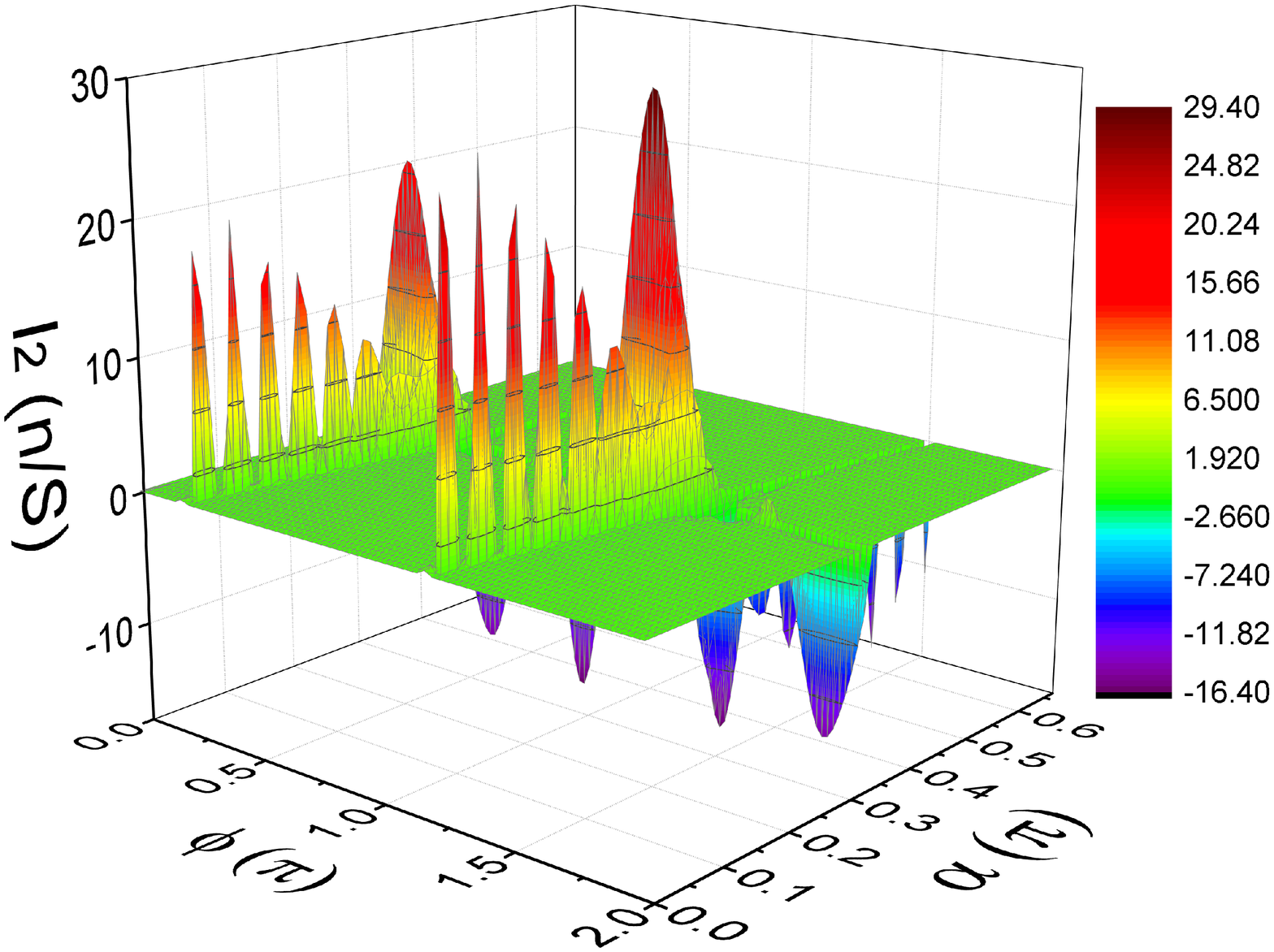}\\
\includegraphics[width=0.24\textwidth, clip]{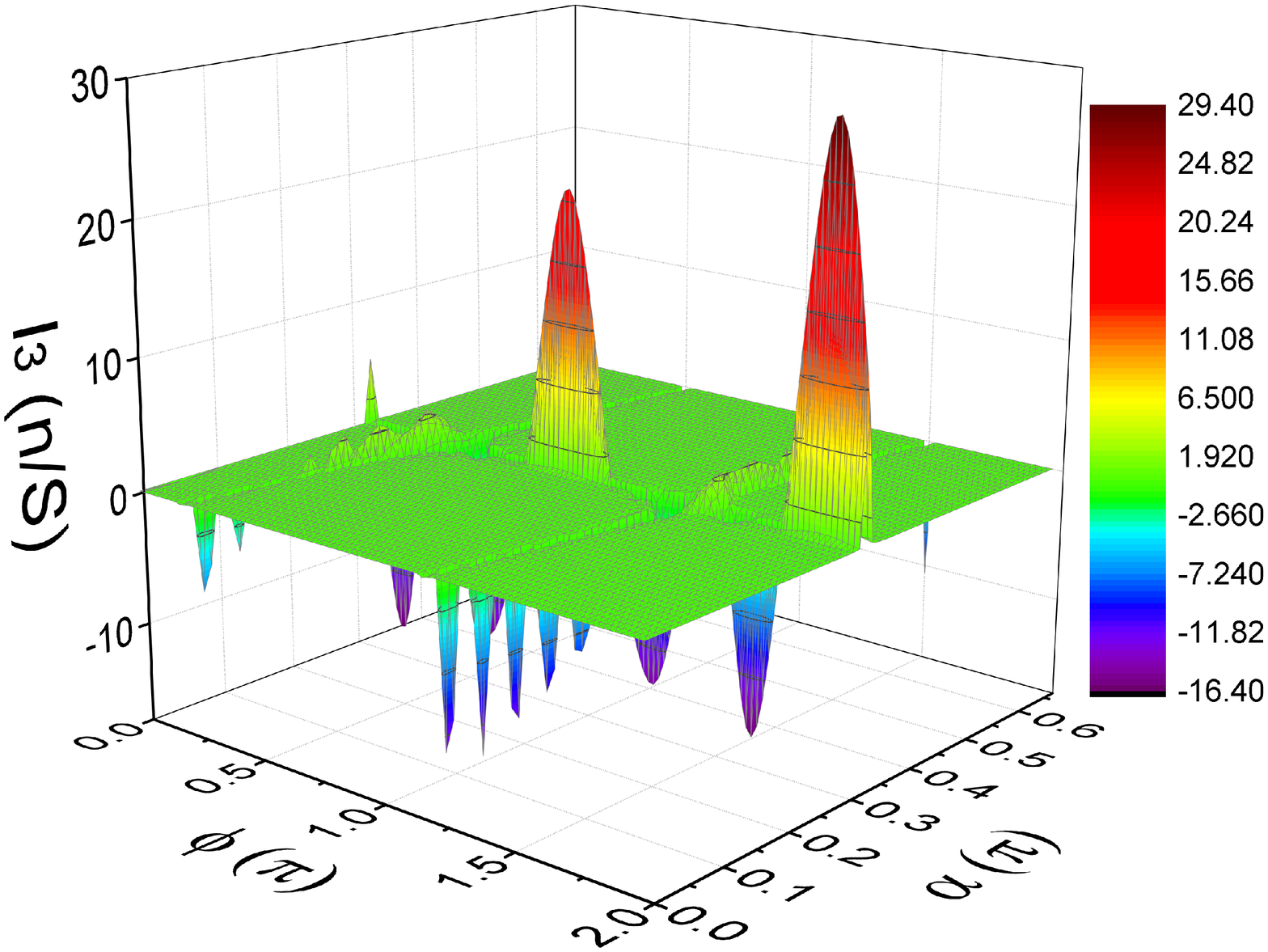}\\
\caption{Atomic current as a function of the laser direction $|\theta|$; (b)Atomic current as a function of the laser
direction $|\theta|$ in the case of one pole of the audion is shut down $(\Gamma_{3}=0)$. Current direction of the
atomic audion corresponding to the laser direction $\theta$ from $0^{\circ}$ to $120^{\circ}$. }%
\label{c1-p-a}%
\end{figure}

The optical traps can be realized in experiment in which a few atoms are bounded\cite{Mancini}. This model should be used for other similar Fermion neutral atoms with long time clock transitions.

In conclusions, atom flow can be created using a artificial gauge field and the direction of current on the three poles can be controlled by tuning the angle of the clock laser. The system of  controllable atom flow in three direction works as an audion and may important for the study of quantum atomic devices.

\begin{acknowledgments}
This work was supported by the National Key R/D Program of China under grants Nos. 2016YFA0301500, NSFC under grants Nos. 11434015, 61227902, 11611530676, SPRPCAS under grants No. XDB01020300, XDB21030300, The NSF of Beijing under Grant No. 1173011, the Scientific Research Project of BMEC under Grant No. KM201711232019, and the Qin Xin Talents Cultivation Program of BISTU under Grant No. QXTCP C201711.
\end{acknowledgments}


\begin{thebibliography}{99}


\bibitem{Dalibard} J. Dalibard, F. Gerbier, G. Juzeli\"{u}nas, and P. \"{O}hberg, Rev. Mod. Phys. \textbf{83}, 1523 (2011).

\bibitem{Livi} L. F. Livi, G. Cappellini, M. Diem, L. Franchi, C. Clivati, M. Frittelli, F. Levi, D. Calonico, J. Catani, M. Inguscio, and L. Fallani, Phys. Rev. Lett. \textbf{117}, 220401 (2016).

\bibitem{Kolkowitz} S. Kolkowitz, S. L. Bromley, T. Bothwell, M. L. Wall, G. E. Marti, A. P. Koller, X. Zhang, A. M. Rey and J. Ye, Nature (London) \textbf{542}, 66 (2017).

\bibitem{Kennedy} C. J. Kennedy, G. A. Siviloglou, H. Miyake, W. C. Burton, and W. Ketterle, Phys. Rev. Lett. \textbf{111}, 225301 (2013).

\bibitem{Kollath} C. Kollath, A. Sheikhan, S. Wolff, F. Brennecke, Phys. Rev. Lett. \textbf{116}, 060401 (2016).

\bibitem{Mancini} M. Mancini, G. Pagano, G. Cappellini, L. Livi, M. Rider, J. Catani, C. Sias, P. Zoller, M. Inguscio, M. Dalmonte, L. Fallani, Science \textbf{349}, 1510 (2015).

\bibitem{An} F. A. An, E. J. Meier, B. Gadway, Science Advances \textbf{3}, 1602685 (2017).

\bibitem{Zheng} W. Zheng and N.R. Cooper, Phys. Rev. Lett. \textbf{117}, 175302 (2016).


\bibitem{Nicholson12} T. L. Nicholson, M. J. Martin, J. R. Williams, B. J. Bloom, M. Bishof, M. D. Swallows, S. L. Campbell, and J. Ye, Phys. Rev. Lett. \textbf{109}, 230801 (2012).

\bibitem{Hinkley} N. Hinkley, J. A. Sherman, N. B. Phillips, M. Schioppo, N. D. Lemke, K. Beloy, M. Pizzocaro, C. W. Oates, A. D. Ludlow, Science \textbf{341}, 1215 (2013).

\bibitem{Bloom} B. J. Bloom, T. L. Nicholson, J. R.Williams, S. L. Campbell, M. Bishof, X. Zhang, W. Zhang, S. L. Bromley, and J. Ye, Nature (London) \textbf{71}, 506 (2014).

\bibitem{Nicholson15} T.L. Nicholson, S.L. Campbell, R.B. Hutson, G.E. Marti, B.J. Bloom,w, R.L. McNally, W. Zhang,
M.D. Barrett, M.S. Safronova, G.F. Strouse, W.L. Tew, and J. Ye, Nature (London) \textbf{6}, 6896 (2015).

\bibitem{Huang} Y. Huang, H. Guan, P. Liu, W. Bian, L. Ma, K. Liang, T. Li, and K. Gao, Phys. Rev. Lett. \textbf{116}, 013001 (2016).

\bibitem{Marti} G. E. Marti, R. B. Hutson, A. Goban, S. L. Campbell, N. Poli, and J. Ye, Phys. Rev. Lett. \textbf{120}, 103201 (2018).



\bibitem{Aidelsburger13} M. Aidelsburger, M. Atala, M. Lohse, J. T. Barreiro, B. Paredes, and I. Bloch, Phys. Rev. Lett. \textbf{111}, 185301 (2013).

\bibitem{Miyake} H. Miyake, G. A. Siviloglou, C. J. Kennedy, W. C. Burton, and W. Ketterle, Phys. Rev. Lett. \textbf{111}, 185302 (2013).

\bibitem{Jotzu} G. Jotzu, M. Messer, R. Desbuquois, M. Lebrat, T. Uehlinger, D. Greif, and T. Esslinger, Nature (London) \textbf{515}, 237 (2014).

\bibitem{Atala} M. Atala, M. Aidelsburger, M. Lohse, J. T. Barreiro, B. Paredes, and I. Bloch, Nature Physics \textbf{10}, 588 (2014).

\bibitem{Aidelsburger15} M. Aidelsburger, M. Lohse, C. Schweizer, M. Atala, J. T. Barreiro, S. Nascimb¨¨ne, N. R. Cooper, I. Bloch, and N. Goldman, Nature Physics \textbf{11}, 162 (2015).


\bibitem{Wall} M. L. Wall, A. P. Koller, S. M. Li, X. B. Zhang, N. R. Cooper, J. Ye, M. Rey, Phys. Rev. Lett. \textbf{116}, 035301 (2016).

\bibitem{Recati} A. Recati, P. O. Fedichev, W. Zwerger, J. von Delft, P. Zoller, Phys. Rev. Lett. \textbf{94}, 040404 (2005).

\bibitem{Bauer} J. Bauer, C. Salomon, and E. Demler, Phys. Rev. Lett. \textbf{111}, 215304 (2013).

\bibitem{Knap} M. Knap, D. A. Abanin, and E. Demler, Phys. Rev. Lett. \textbf{111}, 265302 (2013).




\bibitem{Wenxi} W. Lai, Y. Cao and Z. Ma, J. Phys.: Condens. Matter \textbf{24}, 175301 (2012).



\end{thebibliography}
\end{document}